\documentclass[onecolumn,12pt, draftclsnofoot, margin=1in]{IEEEtran}
\def\withcolors{1}
\def\withnotes{1}
\normalsize
\def\BibTeX{{\rm B\kern-.05em{\sc i\kern-.025em b}\kern-.08em
    T\kern-.1667em\lower.7ex\hbox{E}\kern-.125emX}}

\usepackage[utf8]{inputenc} 
\usepackage{hyperref} 
\usepackage{url} 
\usepackage{booktabs}
\usepackage{amsfonts}
\usepackage{multirow}
\usepackage[noend]{algcompatible}
\usepackage[noend]{algpseudocode}
\usepackage{adjustbox}
\usepackage{soul}
\usepackage{mleftright}
\usepackage{varwidth}
\usepackage[utf8]{inputenc} 
\usepackage[T1]{fontenc} 
\usepackage{hyperref}
\usepackage{url} 

\usepackage[usenames,dvipsnames,table]{xcolor}
\usepackage{amsmath}
\usepackage{algorithm}
\usepackage[noend]{algpseudocode}
\algsetblock[Name]{Encoder}{Stop}{3}{1cm}
\algsetblock[Name]{Decoder}{Stop}{3}{1cm}
\algsetblock[Name]{SharedRandomness}{Stop}{3}{1cm}

\usepackage[noadjust]{cite}
\usepackage{%
    amssymb,%
    amsthm,
    bbm,%
    etoolbox,%
    mathtools,%
    stmaryrd%
}

\usepackage{tikz,pgf,pgfplots,circuitikz}

\theoremstyle{plain} \newtheorem{thm}{Theorem}[section]
\newtheorem{lem}[thm]{Lemma}

\theoremstyle{definition} \newtheorem{defn}[thm]{Definition}

\theoremstyle{remark} \newtheorem{rem}{Remark}

\definecolor{lightgray}{gray}{0.9}

\newcommand{\eqdef}{:=}

\newcommand{\ceil}[1]{\left\lceil#1\right\rceil}%
\newcommand{\norm}[1]{\|#1\|}%
\newcommand\numberthis{\addtocounter{equation}{1}\tag{\theequation}}
 \newcommand{\R}{\mathbb{R}}
\newcommand{\N}{\mathbb{N}}
\newcommand{\bPr}[1]{\mathbb{P}\left(#1\right)}
\newcommand{\E}[1]{\mathbb{E}\left[#1\right]}

\newcommand{\C}{\mathcal{C}}

\newcommand{\X}{\mathcal{X}}
\newcommand{\bX}{\mathbb{X}}

\newcommand{\oO}{\mathcal{O}}
\newcommand{\Q}{\mathcal{Q}}

\newcommand{\uRoman}[1]{\uppercase\expandafter{\romannumeral#1}}

\DeclarePairedDelimiter\floor{\lfloor}{\rfloor}

\usepackage{subcaption}
\DeclareCaptionSubType*{algorithm}

\DeclareCaptionLabelFormat{alglabel}{Alg.~#2}

\ifnum\withcolors=1
  \newcommand{\mcolor}[1]{{\color{ForestGreen}#1}} 
  \newcommand{\tcolor}[1]{{\color{Orange}#1}} 
\else

  \newcommand{\mcolor}[1]{{#1}}
  
  \newcommand{\tcolor}[1]{{#1}}
\fi

\ifnum\withnotes=1
  \newcommand{\mnote}[1]{\par\mcolor{\textbf{M: }\sf #1}} 
  \newcommand{\tnote}[1]{\par\tcolor{\textbf{T: }\sf #1}} 

\else
  \newcommand{\mnote}[1]{}
  \newcommand{\tnote}[1]{}
  
\fi
\newcommand{\ignore}[1]{\leavevmode\unskip} 

\newcommand{\indic}[1]{\mathbbm{1}_{#1}}

\usepackage{soul}

\newcommand{\cE}{\mathcal{E}}
\newcommand{\cO}{\mathcal{O}}
\newcommand{\cC}{\mathcal{C}}
\newcommand{\cN}{\mathcal{N}}

\newcommand{\SNR}{\mathtt{SNR}}
\newcommand{\clientset}{\mathcal{C}}

\input{tangocolors.sty} 

\usepackage{pstricks,pifont}
\usepackage{tikz}
\usepackage{pgfplots,pgfplotstable}
\usepackage{amsmath}

\definecolor{red1}{rgb}{0.4,0,0}
\hypersetup{
colorlinks,
 citecolor=blue,
  linkcolor=red1,
  urlcolor=Blue
}

\usepackage{lipsum}

\allowdisplaybreaks
\begin{document}
\markboth{}
{Draft Version}
\title{Fundamental Limits of Distributed Optimization over  Multiple Access Channel}
\author{Shubham K Jha
        \thanks{
   This work was supported by Prime Minister's Research Fellowship (PMRF), Ministry of Education (MoE), India. A preliminary version of this work \cite{JM23} has appeared in IEEE Information Theory Workshop (ITW), Saint-Malo, France,  2023. \par 
   The author is with Robert Bosch Center for Cyber-Physical Systems, Indian Institute of Science, Bangalore. Email: \url{shubhamkj@iisc.ac.in}.} 
   }
\maketitle 
\begin{abstract}
 We consider distributed optimization over a $d$-dimensional space, where  $K$ remote clients send coded gradient estimates over an {\em additive Gaussian Multiple Access Channel (MAC)} with noise variance $\sigma_z^2$. 
 Furthermore, the codewords from the clients must satisfy the average power constraint $P$, resulting in a signal-to-noise ratio (SNR) of $KP/\sigma_z^2$. 
In this paper, we study the fundamental limits imposed by MAC on the {convergence rate of any distributed optimization algorithm and design optimal communication schemes to achieve these limits.} 
Our first result is a lower bound for the convergence rate, showing that communicating over a MAC imposes a slowdown of $\sqrt{d/\frac{1}{2}\log(1+\SNR)}$ on any protocol compared to the centralized setting. Next, we design a  computationally tractable {digital} communication scheme that matches the lower bound to a logarithmic factor in $K$ when combined with a projected stochastic gradient descent algorithm. At the heart of our communication scheme is carefully combining several compression and modulation ideas such as quantizing along random bases, {\em Wyner-Ziv compression},  {\em modulo-lattice decoding}, and {\em amplitude shift keying.} We also show that analog schemes, which are popular due to their ease of implementation, can give close to optimal convergence rates at low $\SNR$ but experience a slowdown of roughly $\sqrt{d}$ at high $\SNR$.
 \end{abstract}
\date{}
\section{Introduction}
In over-the-air distributed optimization \cite{Amiria19, Amiri19}, the server wants to minimize an unknown function by getting gradient updates from remote clients. In this setting, the clients must communicate their gradient updates over-the-air, namely through a wireless communication channel, to the server. Due to its applications in federated learning \cite{konevcny2016federated}, many interesting schemes have been recently proposed for this problem \cite{sahin2022survey,
Chang20,Sery20, Kobi20,Yang20,  Zhang21, Zhu21,
saha2021decentralized}. However, a clear understanding of the fundamental limits of over-the-air distributed optimization is not present. In this paper, we close this gap by characterizing the fundamental limits imposed on first-order distributed optimization due to over-the-air gradient communication. We also design computationally tractable over-the-air optimization protocols which are almost optimal.

We consider the setting where a server wants to minimize an unknown  smooth convex function with domain in $\R^d$ by  making gradient queries 
to $K$ clients.{ Each of the $K$ clients can generate gradient estimates within a bounded Euclidean distance $\sigma$ of the true gradient.} The clients can communicate their gradient estimates over an {\em additive Gaussian Multiple Access Channel} (MAC) with variance $\sigma_z^2$. Furthermore, each client's communication must also satisfy a power constraint of $P$, which results in a signal-to-noise ratio ($\SNR$) of $KP/\sigma_z^2$. We establish an information-theoretic lower bound on the convergence rate of any over-the-air optimization protocol. Our lower bound shows that there is $\left(\sqrt{\frac{d}{\min(\frac{1}{2}\log(1+\SNR),d)}}\right)$ factor slowdown in convergence rate of any over-the-air optimization protocol when compared to that of centralized setting. Next, we design a digital, computationally tractable communication scheme that, combined with the standard {\em projected stochastic gradient descent} (PSGD) algorithm, almost matches this lower bound. 

We elaborate on several key ideas in our communication scheme. In this scheme, we divide the clients into two halves and send the gradients updates from the first half of the clients to form a preliminary estimate. We then employ {\em Wyner-Ziv} compression to send gradient updates from the second half of clients. This first step is crucial in getting close-to-optimal dependence on the parameter $\sigma$ in the convergence rate. We also employ quantizing along random bases to get optimal dependence on the dimension $d$ in the convergence rate. Finally, to send a $d$-dimensional gradient update in a minimum number of channel uses, we use {\em lattice encoding} and {\em a modulo lattice decoder}, and {\em amplitude shift keying} (ASK) modulation.

We also derive tight lower and upper bounds on the performance of analog schemes. Our bounds show that analog schemes are close to the optimal performing schemes at low $\SNR$, but they are highly suboptimal at high $\SNR$ and have a slowdown of $\sqrt{d}$ as $\SNR$ tends to infinity. Table \ref{T:results} provides a concise summary of all our results.
\begin{table*}
\centering
\caption{Convergence rates of our proposed schemes for large $K$, $N$, and for $\frac{1}{2}\log (1+ \SNR)$ less than $d$.}
\label{T:results}
\renewcommand{\arraystretch}{0.4}
\adjustbox{max width=\linewidth}{
\begin{tabular}{ |c |c |c |c |}
\hline
 & & &\\
Lower Bound & Proposed Scheme & Lower Bound & Proposed Scheme\\
(General) & (General) & (Analog) & (Analog)\\
 & & &\\
$\displaystyle{\frac{D\sigma}{\sqrt{KN}} \cdot \sqrt{\frac{d}{\frac{1}{2}\log (1+ \SNR)}}}$ &$\displaystyle{\frac{D \sqrt{B\sigma}}{\sqrt{KN}} \cdot \sqrt{\frac{d (\log K + \log \log N)}{\frac{1}{2}\log (1+ \SNR)}}}$ &$\displaystyle{\frac{D\sigma}{\sqrt{KN}} \cdot \sqrt{\frac{d}{\SNR}}}$ & $\displaystyle{\frac{D B}{\sqrt{KN}} \cdot \sqrt{\frac{d}{\SNR}}}$\\
& & &\\(Theorem \ref{t:OTAlb}) & (Theorem \ref{t:UWZSGD}) &  (Theorem \ref{t:ALB}) & (Theorem \ref{t:AUBM})\\
 & & &\\
\hline
\end{tabular}}
\end{table*}

Our work is closely related to \cite{JMT22jsait} and \cite{JMST22}. \cite{JMT22jsait}, too, studies fundamental limits of over-the-air optimization, but they do so in the single client setting and when the communication channel is the more straightforward additive Gaussian noise channel. The application of distributed optimization considered in \cite[Section 5]{JMST22} is similar to ours. However, in their setup, the $K$ remote clients can perfectly communicate any update up to $ r$ bits. 
While the more complicated channel considered in this paper prohibits the direct application of schemes from these papers, we build on the ideas proposed in these two papers to come up with our almost optimal scheme.

In a slightly different direction, distributed optimization with compressed gradient estimates has also been extensively studied in recent years (see, for instance,  \cite{alistarh2017qsgd, 
 gandikota2019vqsgd, basu2019qsparse, seide20141, wang2018atomo, wen2017terngrad,     mayekar2020ratq, lin2020achieving, mayekar2020limits,  jhunjhunwala2021adaptive, ghosh2020distributed,  reisizadeh2020fedpaq, suresh2017distributed,  chen2020breaking, prathamesh_information_Constraint_NeurIPS, acharya2019distributed, huang2019optimal, konevcny2018randomized, davies2021new}). Here gradient compression is employed to mitigate the slowdown in convergence when full gradients are communicated. 

The rest of the paper is organized as follows. We setup the problem in the next section and provide basic preliminaries and lower bounds in Section~\ref{s:prelims_lower}.  Section~\ref{s:results} and \ref{s:analog} contain our proposed schemes and the associated results.  All the proofs are given in Section~\ref{s:proof}.  Section~\ref{s:exp} contains the experiments, followed by the concluding remarks in Section~\ref{s:concl}.

\section{Setup}
Consider the following distributed optimization problem. A {\em server} wants to minimize an unknown convex function $f \colon \X \to \R$ over its domain $\X \subset \R^d$ using gradient updates from $K$ remote clients. At each iteration, the server queries the clients for gradient estimates of the unknown function. On receiving the query, each of the $K$ clients generates a stochastic gradient estimate of the function at the queried point, encodes it, and transmits it over a MAC. The output of this channel is available to the server, which it first decodes and then uses it to update the query point for the next iteration using a first-order optimization algorithm (such as Stochastic Gradient Descent). This setting models practical distributed optimization scenarios arising in federated learning and is of independent theoretical interest.

Our goal is twofold: 1) To understand the fundamental limits imposed by communicating gradients over a MAC on the convergence rate; 2) To design the encoding algorithms at the clients, and the decoding and optimization algorithm at the server to come close to the aforementioned fundamental limit.
\subsection{Functions and gradient estimates}
\paragraph{Convex and smooth function family} We assume that the server wants to minimize an unknown function  $f$ which is convex and $L$-smooth functions. That is,  for all\footnote{$\norm{\cdot}$ refers to the standard euclidean norm.} $ x,y\in \X$,
\begin{align}
f(\lambda x+(1-\lambda)y)&\leq \lambda f(x)+(1-\lambda)f(y),\label{e:convex}\\
f(y)-f(x)&\leq \nabla f(x)^{\top}(y-x)+\frac{L}{2}\|y-x\|^2.\label{e:smooth}
\end{align}
\paragraph{Stochastic gradient estimates}
We assume that client $C_k, k\in [K],$ outputs a noisy gradient $\hat{g}_k(x)$ at a query point $x\in \X$ which satisfies
the following standard conditions:
\begin{align}
 \E {\hat{g}_k(x)|x} &= \nabla f(x), \ \text{(unbiasedness)}
\label{e:unbiasedness}
\\
\E{\|\hat{g}_k(x)-\nabla f(x)\|^2|x} &\leq \sigma^2, \ \text{(bounded deviation)}
\label{e:mean_square}\\
\|\hat{g}_k(x)\|^2&\leq B^2. \ \text{(almost surely bounded)}
\label{e:max_norm}
\end{align}

Denote by $\cO$ the set of tuple $(f,\cC)$
of functions and clients satisfying the conditions \eqref{e:convex}, \eqref{e:smooth}, \eqref{e:unbiasedness}, \eqref{e:mean_square} and \eqref{e:max_norm}.

\subsection{{Communication schemes} and the multiple access channel}
For the $t$th query $x_t$ made by the server,  clients $C_1,\dots, C_{K}$ generate gradient estimates $\hat{g}_{1, t},\dots, \hat{g}_{K, t}$, respectively.  In our setting, these gradient estimates are not directly available to the server. 
These are first encoded by the clients for error correction and then transmitted over MAC, and only the output of the channel is available to the server. For all the clients, we consider encoders of length $\ell$ with average power less than $P$. 
That is, the encoder $\varphi_k \colon \R^d \times \mathcal{U} \to \R^\ell$ used by client $C_k$ satisfies the power constraint 
\begin{align}\label{e:Power_constraint}
\E{\|\varphi_k(\hat{g}_{k, t},U)\|^2}\leq \ell P, \quad \forall k \in [K],
\end{align}
where $U$ is string of public randomness available to all the $k$ clients' encoders and the server's decoder, and $\mathcal{U}$ is the space of such random strings. 
For notational convenience, we will drop the argument $U$ of $\varphi_k$ in the rest of the paper.

The encoded codewords  $\{\varphi_k(\hat{g}_{k, t})\}_{k=1}^{K}$ are sent over  MAC using $\ell$ channel uses. 
The server receives the channel output $Y_t\in \R^\ell$ given by
\begin{align}\label{e:Gaussian_channel_M}
Y_{t}(j)=\sum_{k=1}^K \varphi_k(\hat{g}_{k,t})(j)+Z_{t}(j), \qquad
 \forall j \in [\ell],
\end{align} where $Z_{t}(j)$ is Gaussian distributed with mean 0 and variance $\sigma_z^2$. We denote the {\em signal-to-noise ratio} by
$\displaystyle{
\SNR \coloneqq \frac{KP}{\sigma_z^2}
}$.

The decoder $\psi \colon \R^\ell \times \mathcal{U} \to \R^d $ at the server projects back the $\ell$-length channel output to a vector in $\R^d$, which the optimization algorithm uses to update the query point. 

Any tuple of mappings $(\varphi_1, \ldots, \varphi_k, \psi )$, is said to be a {{\em $(d, \ell, P, K)$-communication scheme}} if $\varphi_k$, $k \in [K],$ and $\psi$ are described as above.
Denote by $\mathcal{Q}_{\ell}$ the set of all possible $(d, \ell, P, K)$-communication schemes.
\subsection{Over-the-air optimization over MAC}\label{s:probF}
We now describe the optimization algorithm $\pi$ interacting with  the tuple $(\varphi_1, \ldots, \varphi_k, \psi ) \in  \mathcal{Q}_{\ell}.$ At iteration $t$, the optimization algorithm uses all the previous query points, $\{x_{t^{\prime}}\}_{t^{\prime}=1}^{t-1} $, and the decoded gradient estimates, $\{\psi(Y_{t^{\prime}})\}_{t^{\prime}=1}^{t-1},$   to decide on the query point $x_t \in \X$. The server then queries the clients at the point $x_t$, resulting in a gradient estimate  $\psi(Y_{t})$. This continues for $T$ iterations, after which the algorithm outputs a point $x_T \in \X.$

Denote by $\Pi_{T, \ell}$ the set of all optimization algorithms $\pi$ making $T$ queries to the clients and interacting with a $(d, \ell, P, K)$-{communication scheme.}

For an optimization algorithm $\pi \in \Pi_{T, \ell}$ and a {communication scheme} $Q \in \Q_{\ell,}$ we call the tuple $(\pi, Q)$  an {\em over-the-air  optimization  protocol}. For a tuple of function and clients $(f, \C) \in \oO$, we measure the performance of any over-the-air optimization protocol $(\pi, Q)$ by  the convergence error 
\[\cE(f, \cC, \pi, Q) \coloneqq \E {f(\bar{x}_T)}-\min_{x\in \X}f(x).\]
We will study this error when  the total number of
channel uses, $T\ell$, is restricted 
to be at most $N$.  
We can use {communication schemes} of arbitrary length $\ell$. 
Note, however, that to increase the length $\ell$, we must decrease the number of queries $T$, since the total number of channel uses is limited to $N.$ Conversely, to increase the number of queries, we must decrease the length of the communication schemes. Let $\Lambda(N) \coloneqq  \{(\pi, Q): \pi \in \Pi_{T, \ell}, Q \in \mathcal{Q}_{\ell},  T \ell \leq N\}$
be the set of  over-the-air  optimization protocols with at most $N$ channel uses.
The smallest worst-case error possible over all such protocols is given by
\[\cE^\ast(N, K, \SNR, \X):= \inf_{(\pi, Q) \in \Lambda(N)}\sup_{(f,\cC)\in \cO}
\cE(f, \cC, \pi, Q).\]
Let $\bX :=\{\X\colon \sup_{x,y \in \X}\norm{x-y}\leq D\}.$
In this paper, we will characterize\footnote{While our upper bound techniques can handle an arbitrary, fixed $\X$, the supremum over $\bX$ is to ensure that the lower bounds are independent of the geometry of set $\X$.} $
\cE^\ast(N, K, \SNR):=\sup_{\X \in \bX }\cE^\ast(N, K, \SNR, \X).$





\section{Preliminaries and an information theoretic lower bound}\label{s:prelims_lower}
\subsection{A benchmark from prior results}
We recall the results for the centralized case, which we can model by setting $\SNR=\infty$. In this case, clients can perfectly communicate the gradient estimates in only one channel use. Denote by $\cE^\ast(N, K, \infty)$ the smallest worst-case optimization in this case. A direct application of {\cite[Theorem 6.3]{bubeck2015convex}} leads to the following upper bound on $\mathcal{E}^{\ast}{(N, K, \infty)}$ which serves as a basic benchmark for our results in this paper.
\begin{thm}\label{t:bl}
$
\displaystyle{  \mathcal{E}^{\ast}{(N, K, \infty)}
\leq \frac{\sqrt{2}D\sigma}{\sqrt{KN} } +\frac{LD^2}{2N}}.
$
\end{thm}
\subsection{A general convergence bound}
Throughout the paper, we will use projected stochastic gradient descent (PSGD) as the first-order optimization algorithm $\pi$; the overall over-the-air optimization protocol is described in Algorithm \ref{a:SGD_Q}.  PSGD proceeds as stochastic gradient descent with the additional projection step where it projects the updates back to domain $\X$ using the map $\Gamma_{\X}(y) \coloneqq \min_{x\in \X}\|x-y\|$, $\forall\, y\in \R^d$.
\setcounter{figure}{0}   
\begin{figure}[h]
\centering
\begin{tikzpicture}[scale=1, every node/.style={scale=1}]
\node[draw,text width= 6 cm , text height= ,] {%
\begin{varwidth}{\linewidth}       
            \algrenewcommand\algorithmicindent{0.2em}
\begin{algorithmic}[1]
   \For{$t=0$ to $T-1$}
	\State~~$x_{t+1}{=}\Gamma_{\X} \left(x_{t}-\eta_t \psi(Y_t)\right)$
   \EndFor \State Output $\bar{x}_T=\frac 1 T {\sum_{t=1}^T x_t}$
\end{algorithmic}
\end{varwidth}};
 \end{tikzpicture}
 \renewcommand{\figurename}{Algorithm}
\caption{PSGD for over-the-air optimization}\label{a:SGD_Q}
\end{figure}
The convergence rate of Algorithm \ref{a:SGD_Q} is controlled by the square root of worst-case root mean square error (RMSE) $\alpha(Q)$ 
and the worst-case bias $\beta(Q)$ of the gradient estimates decoded by the server. They are defined as follows:
\begin{align*}
&\alpha(Q)\coloneqq\sup_{\substack{\forall x,k\in [K],\hat{g}_k \in \R^d:\\ \mathbb{E}\|\hat{g}_k-\nabla f(x)\|^2\leq \sigma^2}} \sqrt{\E{\| \psi(Y)-\nabla f(x)\|^2}}, \\
&\beta(Q)\coloneqq \sup_{\substack{\forall x,k\in [K],\hat{g}_k \in \R^d:\\ \mathbb{E}\|\hat{g}_k-\nabla f(x)\|^2\leq \sigma^2}}\|\E{\psi(Y)}-\nabla f(x)\|, 
\end{align*}
where for $i\in [d]$, ${Y}(i)$ satisfies \eqref{e:Gaussian_channel_M} and the expectation is taken over all the randomness in the set up. We now recall a lemma from \cite{JMST22} that upper bounds the convergence rate of Algorithm \ref{a:SGD_Q} in terms of $\alpha(Q)$ and $\beta(Q)$.
\begin{lem}[{\cite[Lemma II.2]{JMST22}}]\label{l:conv}
Let $\pi$ be the PSGD algorithm making $T$ queries to the clients and $Q$ be any communication scheme in $\Q_{\ell}$. Moreover, the over-the-air optimization protocol uses the MAC channel $N=T\cdot \ell$ times. Then, we have $\sup_{(f,O)\in \cO}
\cE(f, \C,\pi, Q)$
\begin{align*}
&\leq \frac{\sqrt{2}D\alpha(Q)}{\sqrt{N/\ell}} +\beta(Q)\left(D+\frac{DB}{\alpha(Q)\sqrt{2N/\ell}}\right)+\frac{LD^2}{2N/\ell}.
\end{align*}
with the learning rate $\eta_t{=}
\min\left\{\frac{1}{L}, \frac{D}{\alpha(Q)\sqrt{2T}}\right\}, \forall t {\in}[T]$. 
\label{thm1} 
\end{lem}
As a result,  it is enough to control the RMSE $\alpha$ and bias $\beta$ of the communication scheme $Q$ to upper bound the overall convergence rate of the corresponding OTA optimization protocol. 

\begin{rem}\label{c5r:DME}
    We remark here that for the noiseless case, i.e., $\SNR=\infty$, the choices for $\varphi$ and $\psi$ are \textit{identity} and \textit{averaging} functions, respectively. Specifically, $\varphi(\hat{g}_{k,t})=\hat{g}_{k,t}$ and $\psi(Y) = Y/K$ with $Y = \sum_k \hat{g}_{k,t}$.
Further, due to no constraint on power, the $d$ coordinates can be sent in just one channel use, $\ell = 1$. That gives $\alpha(Q) = \frac{\sigma}{\sqrt{K}}$ and $\beta(Q) = 0$ retrieving the result in Theorem \ref{t:bl}. 
Therefore, in a nutshell, the primary goal is to design an ``efficient'' distributed mean estimator under MAC constraints in the sense that its performance parameters $\alpha,  \beta$, and $\ell$ are close to these ideal values.
\end{rem}
\subsection{Lower bound for over-the-air optimization}
We now present an information-theoretic lower bound for any over-the-air optimization protocol. We note that 
\cite{JMT22jsait} shows a similar lower bound in the single client setting. We build on their proof and extend the result to the more general setting of $K$ clients.
The key step involves showing that over-the-air optimization over parallel independent additive Gaussian noise channel is much easier than over MAC and then proceeding as in \cite{JMT22jsait}.
\begin{thm}\label{t:OTAlb}
For some universal constant $c\in (0,1)$ and $N\geq \frac{d}{K\log (1+\SNR)}$, we have 
\[\cE^\ast(N,K, \SNR)\geq \frac{cD\sigma}{\sqrt{KN}}\sqrt{\frac{d}{\min\{d, \frac{1}{2}\log (1+\SNR)\}}}.\]
\end{thm}
Our lower bound states  that, except for very high values of $\SNR$, any over-the-air optimization protocol will experience a slowdown  by a factor of  $\sqrt{\frac{d}{ \frac{1}{2}\log (1+\SNR)}}$ over the convergence rate of centralized setting. 
\section{ A digital communication scheme for over-the-air optimization}\label{s:results}
In this section, we present our main result: a digital communication scheme that,  combined with PSGD, will almost match the lower bound in  Theorem \ref{t:OTAlb}. Our scheme below is ``universal'' in the sense that the clients don’t
require the knowledge of $\sigma$ for the transmission of gradient estimates.  As pointed out in Remark \ref{c5r:DME}, our focus should be on designing an efficient distributed mean estimator.
\subsection{Warm-up scheme: {\tt UQ-OTA}}\label{s:warmup}
For ease of presentation, we first present a warm-up OTA optimization protocol, {\tt UQ-OTA}, based on uniform quantization. We will build on the components described below to present our final digital scheme. Throughout the description of our schemes, we omit the subscript $t$ for convenience.
\paragraph{Uniform quantization.}
 Each client $C_k$ first divides the gradient estimate $\hat{g}_{k}$ by the number of clients $K$ to form $\tilde{g}_{k}$ and quantizes it using an unbiased $v$-level coordinate-wise uniform quantizer $v$\texttt{-CUQ}.
The $v$\texttt{-CUQ} takes $i$th coordinate $\tilde{g}_{k}(i){\in} \left[-\frac{B}{K},\frac{B}{K}\right]$ as input and outputs $z_{k,i}\in \{0,...,v-1\}$ as per the following rule:
\begin{align*}
z_{k,i}= 
\begin{cases}  
\ceil{\frac{(v-1)(K\tilde{g}_{k}(i)+B)}{2B}}, \text{ w.p.}  \frac{\tilde{g}_{k}(i) - \floor{\frac{\tilde{g}_{k}(i)K(v-1)}{2B}}}{2B/(K(v-1))}\vspace{0.3cm}\\
 \left\lfloor\frac{(v-1)(K\tilde{g}_{k}(i)+B)}{2B}\right\rfloor, \text{ w.p. } \frac{\ceil{\frac{\tilde{g}_{k}(i)K(v-1)}{2B}} - \tilde{g}_{k}(i)}{2B/(K(v-1))}
\end{cases}.
\end{align*}
That is, the quantizer first finds the two consecutive quantization points containing $\tilde{g}_k(i)$ and declares exactly one of the corresponding indices stochastically. The probability distribution is chosen in such a way that the output $z_{k, i}$ suffices to form an unbiased estimate of $\tilde{g}_{k}(i)$. Define $\mathtt{Z}_k\coloneqq \{z_{k,i}:i\in [d]\}$ as the quantized output for client $k$. We now process the quantized output for transmission over MAC.
\paragraph{Lattice encoding and ASK modulation using $\mathcal{M}({\tt Q}_{k},v,p)$.}
Client $C_k$ sends ${\tt Z}_k$ over MAC  by first encoding them onto a one-dimensional lattice and further modulating them onto an ASK code. 
We describe below the entire procedure and refer to it by $\mathcal{M}({\tt Q}_{k},v,p)$ with parameters $v$ and $p$ to be specified later. 

For an integer\footnote{For simplicity, we assume $p$ divides $d$.} $p\leq d$, we first partition the set of coordinates $[d]$ equally into blocks of size $p$. 
That way, we have $d/p$ blocks. For ${j\in [d/p]}$, denote by $\mathcal{B}_j$ the $j$th block is given by $ \mathcal{B}_j = \{(j-1)p+1, \dots, (j-1)p+p\}.$ For each $\mathcal{B}_j$, the corresponding quantized values are mapped onto a one-dimensional lattice $\Lambda_w$ generated by a set of basis $\{w^0,...,w^{p-1}\}$ for some positive integer $w$ and is given by
\begin{align*}
    \Lambda_w = \{q_1\cdot w^0+\dots+q_p\cdot w^{p-1}: 0\leq q_1,...,q_p\leq v-1\}.
\end{align*}
Denote by $\tau_{k,j}$ the lattice point corresponding to block $\mathcal{B}_j$ for the client $C_k$ is given by
\begin{align*}
    \tau_{k,j}= {\tt Z}_k(\mathcal{B}_j(1))+{\tt Z}_k(\mathcal{B}_j(2))\cdot w+\dots+{\tt Z}_k(\mathcal{B}_j(p)) \cdot w^{p-1},
\end{align*} with $w=K(v-1)+1$. Note that this choice of $w$ ensures successful recovery of the sum of client updates at the server.

\begin{defn}\label{d:ASK_codes}
  A code is an
\emph{Amplitude Shift Keying (ASK)
  code} satisfying the average power constraint \eqref{e:Power_constraint} if the range $\mathcal{A}$
of the encoder mapping is given by
\begin{align*}
\mathcal{A}\coloneqq \left\{-\sqrt{P} +(i-1)\cdot \frac{2\sqrt{P}}{r-1}\colon i \in [r] \right\},
\end{align*}
for some $r\in \N$.  Note that this is a code of length $1$.Note that
this is a code of length $1$.
\end{defn}
To satisfy the power constraints of MAC, we then modulate each $\tau_{k,j}$ to $[-\sqrt{P}, \sqrt{P}]$ using an ASK code. 
Since each $\tau_{k,j}$ takes values in $\{0,...,\frac{w^{p}-1}{K}\}$, we set the size of ASK code  $r=\frac{w^p-1}{K}+1$  to establish one-to-one correspondence.
Consequently, the encoded value is given by 
$
\varphi_k(j)=\mathcal{A}(\tau_{k,j}+1), \forall j\in [d/p], \forall k\in [K].
$
The transmission takes place over MAC as in \eqref{e:Gaussian_channel_M}. 
\paragraph{Lattice decoding at server $\mathcal{L}(Y,v,p)$.} 
On the server side, our goal will be to compute an unbiased estimate of sum $\sum_k \tilde{g}_k$ from $Y$. 
Note that the natural aggregation property of the MAC channel makes it somewhat easier to recover this sum instead of individual ${\tt Z}_k$s.
This further implies recovering the sum of quantized outputs $\sum_k{\tt Z}_{k}$ which suffices to form the unbiased estimate of $\sum_k \tilde{g}_k$.

Towards that, each coordinate $Y(j), j\in [\ell]$, is first fed into a coordinate-wise MD decoder to locate the nearest ASK codeword $\hat{Y}(j)$ in $\{-K\sqrt{P}+(i-1)\cdot \frac{2\sqrt{P}}{r-1}:i\in [r]\}$.
Using the one-to-one correspondence, the decoded point $\hat{Y}(j)$ is then mapped back to the lattice\footnote{Note that under perfect decoding,  this yields the sum of transmitted lattice points $\sum_k\tau_{k,j}$.} $\Lambda_w$. 
Denote by $\hat{\bar{\tau}}_j$ the decoded lattice point
can be expressed as  
\[\hat{\bar{\tau}}_j=\lambda(\mathcal{B}_j(1))
\cdot w^0+\dots+\lambda(\mathcal{B}_j(p))\cdot w^{p-1},\] for some vector $\lambda\in \{0,...,K(v-1)\}^d$. Therefore, to recover the desired sum, the server uses a {\em modulo-lattice decoder} for each $\mathcal{B}_j, j\in [d/p]$, that successively outputs the corresponding coordinates of $\lambda$. In particular, $\forall i \in [p]$,
\begin{align*}
\lambda(\mathcal{B}_j(i))=\frac{\hat{\bar{\tau}}_j-\lambda(\mathcal{B}_j(1))\dots-\lambda(\mathcal{B}_j(i-1))w^{i-2}}{w^{i-1}}\mod w,
\end{align*} where for positive integers $a,b,m$ and $t$ such that $a=m\cdot b+t$ and $0\leq m\leq b-1$, the  mod-operation is defined as ${a\text{ mod }b}=t$. 
Note that such recovery is feasible with the current choice of $w$ as each coordinate of $\sum_{k \in [K]}{\tt Z}_{k}$ is at most $w-1$.
The vector $\lambda$ obtained above is finally used to form $\psi(Y)$ to be used in Algorithm \ref{a:SGD_Q} as
\begin{align}\label{e:est_CUQ}
\psi(Y)=-B+\frac{2B}{K(v-1)}\cdot \lambda.
\end{align}
\begin{thm}\label{t:CUQ}
Let $\pi$ be the optimization algorithm described in Algorithm \ref{a:SGD_Q}, where $\psi(Y)$ is obtained in \eqref{e:est_CUQ} with $v=\sqrt{d}+1$. 
Then,
  for a universal constant $c_1>0$ and integers $p,K$ such that ${d \geq p \geq 1}$ and $K\geq B^2/\sigma^2$, we have
\begin{align*}
\sup_{(f,\cC)\in \cO}\cE(f, \clientset, \pi, Q) &\leq  \frac{c_1DB}{\sqrt{KN}}\sqrt{\frac{d}{p}}+\frac{LD^2d}{2Np},
\end{align*}
where $p=\left\lfloor\frac{\log\left(1+\sqrt{\frac{2K\SNR}{\ln(KN^{1.5})}}\right)}{\log (Kd)}\right\rfloor$.
\end{thm}

\begin{rem}
We remark that both the encoding and decoding complexity of {\tt UQ-OTA} is $O(d)$.
\end{rem}

\begin{rem}\label{r:1}
 We remark that the size of $\mathcal{A}$ used for MAC transmission grows with the operating $\SNR$ as $r\approx \min (\sqrt{d}+1,\floor{1+\sqrt{2\SNR/(K\ln(KN^{1.5}))}})$.
\end{rem}
\begin{rem}\label{rem:uqota}
At large values of $\SNR\geq 2(2^d-1)$, $p\approx \frac{\frac{1}{2}\log(1+\SNR)}{\log(Kd)}$.
We remark that {\tt UQ-OTA} incur a $(B/\sigma)\sqrt{\log K+\log d+\log\log N)}$ factor slowdown in convergence rate compared to that of centralized setting, which can still cause a slowdown for high-dimensional settings and large values of $B$ compared to $\sigma$.
\end{rem}
\subsection{Wyner-Ziv digital scheme: { \tt WZ-OTA}} 
{We are now ready to present our main digital scheme {\tt WZ-OTA} which significantly improves over the performance of {\tt UQ-OTA} and is almost optimal.}

In this scheme, we partition the clients $\cC$ equally into two sets $\cC_1$ and $\cC_2$.
In each iteration $t$, the clients in $\cC_1$ construct the side information at the server, and the remaining clients in $\cC_2$ exploit this information to form a Wyner-Ziv estimate of $\nabla f(x_t)$ at the server.
\paragraph{Side information construction}
The clients in $\cC_1$ use the previously described {\tt UQ-OTA}  communication scheme to form a preliminary estimate \eqref{e:est_CUQ} at the server. This requires $\ell=d/p$ channel uses. Note that the clients in $\cC_2$ send $0$ during these transmissions. 

The server divides this preliminary estimate by $K/2$ to form $S$ and  then rotates it by a random matrix $\mathbf{R}$ to form the side information $\mathbf{R}S$.
 Here $\mathbf{R}=1/\sqrt{d}\mathbf{H}\mathbf{D}^\prime$
where $\mathbf{H}$ is the Walsh-Hadamard\footnote{Without loss of generality, we assume $d$ is a power of 2.  If not, we can zero-pad the gradient estimates and make the resulting dimension power of $2$; this only adds a constant multiplicative factor to our upper bounds.} matrix \cite{horadam2012hadamard}, and $\mathbf{D^\prime}$ is a random diagonal matrix with non-zero entries generated uniformly and independently from $\{-1,+1\}$.
\paragraph{The Wyner-Ziv estimate}
The clients in $\cC_2$ use a Wyner-Ziv estimator boosted DAQ from \cite{JMST22} to construct the final estimate, while those in $\cC_1$ tranmit $0$ in all channel uses.{ The boosted DAQ uses the idea of correlated sampling between the input and the side information to reduce quantization error.}
{Specifically, for an input $|x|\leq M$ at the encoder and a corresponding side information $|y|\leq M$ at the decoder, the boosted DAQ estimate is given by 
\begin{align}\label{e:daq}
\hat{X}=(2M/I)\sum_{i\in [I]}\left(\mathbbm{1}_{\{U_i\leq x\}}-\mathbbm{1}_{\{U_i\leq y\}}\right)+y, 
\end{align}
where each $U_i\sim \mathtt{unif}[-M,M]$ is a uniform random variable.} {Note that $\hat{X}$ is an unbiased estimate of $x$ with MSE at  most $2M|x-y|/I$.}

 In our setting, each client ${C_k\in \cC_2}$ first pre-processes its noisy estimate as 
 $\tilde{g}_{k}=\frac{2\hat{g}_{k}}{K}$ and uses shared randomness to draw $I$ uniform random vectors $U_{k,i}\in [-M, M]^d, i\in [I]$, independently. The choice of $M$ and $I$ are crucial for our scheme and will be specified later. Using shared randomness again, each $\tilde{g}_{k}$ is rotated using the same random matrix $\mathbf{R}$ used earlier.  Each coordinate of this rotated vector 
is then quantized to an element in $\{0,...,I\}$ as
\begin{align*}
{\tt Q}_{k}(j)=\sum_{i\in[I]}\mathbbm{1}_{\{U_{k,i}(j)\leq \mathbf{R}\tilde{g}_{k}(j)\}}, \ \forall j\in [d].    
\end{align*}
{As an aside, it is instructive to note that under the event $\mathcal{V}_j=\{|\mathbf{R}S(j)|\leq M, |\mathbf{R}\tilde{g}_k(j)|\leq M\}$,  ${\tt Q}_{k}(j)$ suffices to form an unbiased estimate of $\mathbf{R}\tilde{g}_{k}(j)$ using boosted DAQ (see \eqref{e:daq}). Coming back to our scheme, each client $k$ transmits the quantized vector ${\tt Q}_{k}$ over the MAC channel by first using the lattice encoder and then using ASK modulation. The entire operation is described by the function
$\mathcal{M}({\tt Q}_{k},v^\prime,p^\prime)$ (see Section \ref{s:warmup}) with $v^\prime=I+1$ and $p^\prime$ to be specified shortly.
Note that there are $\ell=d/p^\prime$ channel uses per iteration.

At the server, the channel output $Y\in \R^{d/p^\prime}$ is passed through $\mathcal{L}(Y, v^\prime, p^\prime)$ to obtain $\lambda$.
Following the boosted DAQ estimator \eqref{e:daq}, 
the final output $\psi(Y)$ is given by
\begin{align}\label{e:wz_output}
\psi(Y)&=\frac{2M}{I}\mathbf{R}^{-1}\sum_{j\in [d]}\left(\lambda(j)-\omega(j)\right)\cdot e_j+\frac{K}{2}
S,
\end{align} 
where each $\omega(j)=\sum_{k\in \cC_2}\sum_{i\in [I]}\mathbbm{1}_{\{U_{k,i}(j)\leq \mathbf{R}S(j)\}}$ can be realized at the server using shared randomness and the available side-information. 
We next characterise the performance of {\tt WZ-OTA}.}
\begin{thm}\label{t:UWZSGD} Let $c_2,c_3$ be positive universal constants and $\pi$ be the optimization algorithm described in Algorithm \ref{a:SGD_Q}, where $\psi(Y)$ is obtained using \eqref{e:wz_output} with $v=7, M=\frac{c_2B}{K\sqrt{d}}\sqrt{\ln(K^{1.5}N)}$ and $I=c_2\sqrt{\ln(K^{1.5}N)}$. Then,
  for integers $p,p^\prime$ and $K$ such that 
 $K\geq B^2d/\sigma^2$ and ${d \geq p,p^\prime \geq 1}$, we have
\begin{align*}
\sup_{(f,\cC)\in \cO}\cE(f, \clientset, \pi, Q) &\leq  \frac{c_3D\sqrt{B\sigma}}{\sqrt{KN}}\sqrt{\frac{d}{q}}+ \frac{LD^2d}{2Nq},
\end{align*}
where $\frac{1}{q} = \frac{1}{p}+\frac{1}{p^\prime}$ with $p=\left\lfloor\frac{\log\left(1+ \sqrt{\frac{K\SNR}{2\ln(KN^{1.5})}}\right)}{\log K}\right\rfloor$ and $p^\prime = \left\lfloor\frac{\log\left( 1+\sqrt{\frac{K\SNR}{2\ln(KN^{1.5})}}\right)}{\log K+\log\log N}\right\rfloor$. 
\end{thm}
\begin{rem}\label{r:wz_better}
    {For large $K, N$, we remark that the {\tt WZ-OTA}  combined with PSGD is off only by a factor of $\sqrt{(B/\sigma) \left(\log K+\log \log N\right)}$ from our lower bound. 
    In comparison, from Theorem \ref{t:CUQ}, {\tt UQ-OTA} combined with PSGD is off by a factor $\left(B/\sigma\right) \sqrt{\left(\log K + \log d +\log\log N\right)}.$ Quantization along  random bases and Wyner-ziv  compression allows {\tt WZ-OTA}  to improve by factors $\log d$  and $\sqrt{B/\sigma}$ over {\tt UQ-OTA}.}
\end{rem}
\begin{rem}
We remark that the random rotation step using the Walsh-Hadamard matrix can be performed in nearly linear-time and, in particular, requires $O(d\log d)$ operations. Since this is the most expensive step in {\tt WZ-OTA}, the encoding and decoding complexity of {\tt WZ-OTA} is $O(d\log d)$.
\end{rem}

\section{Performance of Analog Schemes}\label{s:analog}
\begin{defn}\label{d:analog_schemes}
A {communication scheme} is an \emph{analog scheme} if the encoder mapping $\varphi$ is
  linear, i.e., $\varphi(x)=\mathbf{A}x$ for $\mathbf{A}\in \R^{\ell\times d}$ and $\ell \leq d$.
 We allow random entries for $\mathbf{A}$ as long as the randomness is independent of $x$. 
 For the class of {$(d, \ell, P, K)$-{communication schemes}} restricted to using such analog schemes, we denote by $\cE^\ast_{\tt {analog}}(N,K, \SNR)$ the corresponding min-max optimization error. Clearly,  $\cE^\ast_{\tt analog}(N,K, \SNR) \geq \cE^\ast(N,K, \SNR)$.
\end{defn}
We begin by proving a lower bound for analog communication schemes.
\begin{thm}\label{t:ALB}
For some universal constant $c \in (0, 1), $  and $N \geq \frac{d}{K}(\sigma^2+\frac{\sigma^2}{\SNR})$, we have 
\begin{align*}
\cE^\ast_{\tt analog}(N,K, \SNR) \geq \frac{cD}{\sqrt{KN}}\sqrt{d\sigma^2+\frac{d\sigma^2}{\SNR}}.
\end{align*}
\end{thm}
The following lower bound also uses affine functions as difficult functions and builds on a class of Gaussian oracles proposed, recently, towards proving a similar result in \cite{JMT22jsait}. 

{For our upper bound},  we use the well-known {\em scaled transmission} scheme from~\cite{Kobi20}. In this scheme, the gradient estimates are scaled-down  by $\sqrt{dP}/B$ by every client $C_k \in \C$ to satisfy the power constraint in \eqref{e:Power_constraint}, sent coordinate-by-coordinate over $d$ channel uses, and then scaled-up by $B/\sqrt{dP}$ and averaged at the server before using
it in a gradient descent procedure. It is not difficult to see the following upper bound.
\begin{thm}\label{t:AUBM}
Let $\pi$ be the PSGD optimization algorithm and $Q$ be the scaled transmission communication scheme described above. Then, we have
  \begin{align*}
   \sup_{(f,\cC)\in \cO}
  \cE(f, \cC, \pi, Q)  \leq \frac{\sqrt{2}D}{\sqrt{KN }} \sqrt{d\sigma^2+\frac{dB^2}{\SNR}}+\frac{dLD^2}{2N}.
  \end{align*}
\end{thm}

\begin{rem}\label{r:analog}
For $\SNR\geq B^2/\sigma^2$, Theorem \ref{t:ALB} shows that compared to the centralized setting discussed in   Theorem \ref{t:bl}, analog schemes will have a slowdown of $\sqrt{d}$. However, for small values of $\SNR$,  an analog communication scheme combined with PSGD gives close optimal performance. It matches the lower bound in Theorem \ref{t:OTAlb} up to a factor of $B/\sigma$. This observation follows by noting that $\log(1+\SNR) \approx \SNR$ for small values of $\SNR.$
\end{rem}
\section{Proofs}\label{s:proof}
\subsection*{Difficult functions for lower bound}
For our lower bounds, we use affine functions as difficult functions which are $0$-smooth and are admissible in the class of $L$-smooth functions. These functions are considered in the same spirit as in showing the lower bounds for convex, smooth optimization under communication constraints \cite{JMST22}. 
We consider the domain $\X=\{x \in \R^d:\norm{x}_\infty\leq D/(2\sqrt{d} \}$, and consider the following class of functions on $\X$: For $v \in \{-1, 1 \}^d$, let
\begin{equation*}
  f_{v}(x) \eqdef \frac{2\sigma\delta}{\sqrt{d}} \sum_{i=1}^{d} \mleft|x(i) - \frac{v(i) D}{2 \sqrt{d}}\mright|, \quad \forall\, x\in \X,
\end{equation*}
and $x_v^\ast$ be its minimizer.
Note that the gradient of $f_v$ at $x\in \X$ is independent of $x$, i.e., $-2\sigma\delta v /\sqrt{d}$.  
For any such $f_v$, each coordinate of $d$-dimensional noisy gradient $\hat{g}_{k,t}(i)$ takes $-\sigma/\sqrt{d}$ or $\sigma/\sqrt{d}$ independently with probabilities $(1+2\delta v(i))/2$ and $(1-2\delta v(i))/2$, respectively.
The parameter $\delta >0$ is to be chosen later.
Note that the above construction satisfies the set of assumptions in \eqref{e:unbiasedness}, 
\eqref{e:mean_square} and \eqref{e:max_norm}.
\subsection{Proof of Theorem \ref{t:OTAlb}}
Draw ${V \sim \texttt{unif}\{-1, 1\}^d}$. With respect to the associated random function $f_{V}$, each client $C_i$ chooses a quantizer $\varphi$ to generate output 
$\varphi(\hat{g}_{k,t}).$
Denote by $Y^{T}{=}(Y_{1}, \ldots, Y_{T})$ the vector of $\ell$-dimensional MAC channel outputs observed at the server. 
{We follow a standard approach to reduce the underlying optimization problem to a multiple hypothesis problem of estimating $V$ from output $Y^T$.
}
Denote by $\mathbf{p}_{+i}^{Y^{T}}$ and $\mathbf{p}_{-i}^{Y^{T}}$ the distribution of $Y^{T}$ given $V(i)=+1$ and $V(i)=-1$, respectively.
We can relate the expected gap to optimality to the average total variational distance between $\mathbf{p}_{+i}^{Y^{T}}$ and $\mathbf{p}_{-i}^{Y^{T}}$
using the techniques from 
\cite[Lemma 3, 4]{ACMT21informationconstrained}, which in turn builds on \cite{agarwal2009information, acharya2020general}. In particular, 
\begin{align*} 
 \E{f_{V}(\bar{x}_T)-f_{V}(x_V^\ast)} &= \sum_{i=1}^d  \E{\frac{2\sigma\delta}{\sqrt{d}}\mleft|\bar{x}_T(i) - \frac{V(i) D}{2 \sqrt{d}}\mright|} \\
  & \geq \frac{D\sigma\delta}{3d}\sum_{i=1}^{d} \mathbb{P}\left(\frac{2\sigma\delta}{\sqrt{d}} \mleft|x(i) - \frac{V(i) D}{2 \sqrt{d}}\mright|\geq \frac{D\sigma\delta}{3d}\right)\\
  &\geq  {\frac{D\sigma\delta}{6}}\Bigg[1 - \frac{1}{d}
      \sum_{i=1}^d \mathtt{d_{TV}}\left(\mathbf{p}_{+i}^{Y^{T}}, \mathbf{p}_{-i}^{Y^{T}}\right)\Bigg],
\end{align*}
where the first inequality is Markov's inequality and the second is the standard lower bound on probability of error in binary hypothesis testing under uniform prior.
The 
average total-variational distance term can be further bounded using the convenient ``plug-and-play'' bound from \cite[Theorem 2]{acharya2020general}
\begin{align*}
\left(\frac{1}{d}\sum_{i=1}^d \mathtt{d_{TV}}\left(\mathbf{p}_{+i}^{Y^{T}}, \mathbf{p}_{-i}^{Y^{T}}\right)\right)^2
&\leq \frac{284T\delta^2}{d}
\max_{v\in \{-1,1\}^d} \max_{\varphi\in \mathcal{Q}}
I\left(\hat{g}_{1,1},...,\hat{g}_{K,1}\wedge Y_1\right)\\
&\leq \frac{284T \delta^2}{d}
\cdot \min(\max_{v\in \{-1,1\}^d} \max_{\varphi\in \mathcal{Q}}
I\left(\varphi(\hat{g}_{1,1}), ..., \varphi(\hat{g}_{K,1})\wedge Y_1\right), Kd\ell),
\end{align*}
where the first inequality holds for ${\delta\in (0, 1/6)}$, and the second inequality uses $I\left(\hat{g}_{1,1},...,\hat{g}_{K,1}\wedge Y_1\right)\leq H(I\left(\hat{g}_{1,1},...,\hat{g}_{K,1}\wedge Y_1\right))\leq Kd\ell$ and data-processing for the second. 
Further, we consider an auxiliary generative model for obtaining the MAC output $Y_1$.
Specifically, we assume $K$ parallel Gaussian noise outputs given by
\begin{align*}
    Y_{k,1}(i) = \varphi(\hat{g}_{k,1})(i) + Z_{k,1}(i), \ i\in [\ell], k\in [K],
\end{align*}
where $Z_{k,1}\sim \cN(0,\sigma_z^2/K\mathbf{I}_\ell)$. Note that the output $Y_1$ and the sum of $\sum_{k\in [K]}Y_{k,1}$ are statistically equivalent.
Thus, we can write
\begin{align*}
I\left(\varphi(\hat{g}_{1,1}), ..., \varphi(\hat{g}_{K,1})\wedge Y_1\right) &= I\left(\varphi(\hat{g}_{1,1}), ..., \varphi(\hat{g}_{K,1})\wedge Y_{1,1}+...+Y_{K,1}\right)\\
    &\leq I\left(\varphi(\hat{g}_{1,1}), ..., \varphi(\hat{g}_{K,1})\wedge Y_{1,1},...,Y_{K,1}\right)\\
    &\leq \frac{K\ell}{2}\log (1+\SNR).
\end{align*}
Combining, we have for some universal constant $c^\prime$, 
\begin{align*}
  \E{f_{V}(\bar{x}_T)-f_{V}(x_V^\ast)}&\geq  {\frac{D\sigma\delta}{3}}\Bigg[1 - \sqrt{\frac{c^{\prime}KN\delta^2\min (d,\frac{1}{2}\log (1+\SNR))}{d}}
\Bigg].
\end{align*}
Setting $\delta=\sqrt{d/(2c^{\prime}\min(2d, \log(1+\SNR))KN)},$ we finally get for some universal constant $c\in (0,1)$.
\[\E{f_{V}(\bar{x}_T)-f_{V}(x_V^\ast)}\geq \frac{cD\sigma}{\sqrt{KN}}\sqrt{\frac{d}{\min(d,\frac{1}{2}\log(1+\SNR))}},\]
where we need $N\geq 18d/(c^{\prime\prime\prime}K\log(1+\SNR))$ in order to enforce $\delta\leq 1/6.$  
The proof is completed by noting that
  $\cE^\ast(N,K, \SNR) \geq  \E{f_{V}(\bar{x}_T)-f_{V}(x_V^\ast)}.$
\subsection{A general recipe for upper bounds}
We now provide the general recipe to prove our upper bounds.
For the minimum-distance decoder,  denote by $A_N$ the event where all the ASK constellation points
sent in $N$ channel uses
are decoded correctly by the algorithm and by $A_N^c$ its complement, i.e.,
$A_N^c := \cup_{t=1}^{N} \{|Z_{t}(1)| \geq 2\sqrt{P}/(r-1)\}$,
where $Z_{t}(1)$ is defined in \eqref{e:Gaussian_channel_M}.
We have 
\begin{align*}
\E{f(\bar{x}_T)-f(x^\ast )}&=\E{(f(\bar{x}_T)-f(x^\ast))\mid A_{N}}\cdot \bPr{A_{N}} +\E{(f(\bar{x}_T)-f(x^\ast))\mid A_{N}^c}\cdot \bPr{A_{N}^c}\\
&\leq \E{(f(\bar{x}_T)-f(x^\ast))\mid A_{N}}+DB\cdot \bPr{A_{N}^c}. 
\end{align*}
Using Chernoff's bound: $
\bPr{A_N^c} \leq N\exp\left( -\frac{2K\SNR}{(w^p-1)^2}\right)\leq \frac{1}{K\sqrt{N}},
$
where the last line follows by setting $p=\floor{\log_w\left( \sqrt{\frac{2K\SNR}{\ln(KN^{1.5})}}+1\right)}$. The first term under perfect decoding is bounded by calculating the performance measures ${\alpha(Q)}$ and $\beta(Q)$ for different schemes, and the proof is completed using Lemma \ref{l:conv}.
\subsection{Proof of Theorem \ref{t:CUQ}}
Note that the number of channel uses $\ell=d/p$ per iteration. 
Under perfect decoding, $\lambda=\sum_{k}{\tt Q}_{k}$ implying $\beta(Q)=0$. Using conditional expectation, we also have
\begin{align*}
    \E{\left\|\psi(Y_t)-\sum_{k\in [K]}\tilde{g}_{k,t}\right\|^2}\leq \frac{4B^2d}{K(v-1)^2}.
\end{align*}
Further, from \eqref{e:mean_square} we have $\E{\left\|\sum_{k\in [K]}\tilde{g}_{k,t}-\nabla f(x_t)\right\|^2}\leq \frac{\sigma^2}{K}.$ 
At last, using the inequality $(a+b)^2\leq 2(a^2+b^2)$,
setting $v=\sqrt{d}+1$,  and the fact $\sigma\leq B$,
we get $\alpha^2(Q)\leq \frac{10B^2}{K}$.
\vspace{-0.5cm}
\subsection{Proof of Theorem \ref{t:UWZSGD}}
Based on the proof for \cite[Lemma 4.1]{mayekar2021wyner}, we begin by a lemma capturing the performance of boosted DAQ estimator.
\begin{lem}\label{l:daq}
Given that $x,y\in [-M,M]^d.$ For the boosted DAQ estimate $\hat{X}$ in  \eqref{e:daq}, we have
\begin{align*}
\mathbb{E}[\hat{X}]=x \text{ and }\E{\|\hat{X}-x\|^2}\leq (2M/I)\sqrt{d}\|x-y\|.
\end{align*}
\end{lem}
\noindent We have $\lambda(j){=}\sum_{k\in \cC_2}\sum_{i\in [I]} \mathbbm{1}_{\{U_{k,i}(j)\leq \mathbf{R}\tilde{g}_{k}(j)\}}$ under perfect decoding, which implies
\begin{align*}
\mathbf{R}\psi&=(2M/I)\sum_{j\in [d]}\sum_{k\in \cC_2}\sum_{i\in [I]}\mathbbm{1}_{\{U_{k,i}(j)\leq \mathbf{R}\tilde{g}_{k}(j)\}}e_j-\mathbbm{1}_{\{U_{k,i}(j)\leq \mathbf{R}S(j)\}}e_j+(K/2)\mathbf{R}S.
\end{align*}
Similar to the proof of Theorem \ref{t:CUQ}, we first bound 
$\E{\|\psi(Y)-\sum_{k\in \cC_2}\tilde{g}_{k}\|^2}$ using conditional expectation.
Define 
$
\psi_k{\coloneqq}(2M/I)\mathbf{R}^{-1}\sum_{j\in [d]}\sum_{i\in [I]}(\mathbbm{1}_{\{U_{k,i}(j)\leq \mathbf{R}\tilde{g}_{k}(j)\}}-\mathbbm{1}_{\{U_{k,i}(j)\leq \mathbf{R}S(j)\}})e_j+S,
$
such that $\psi{=}\sum_k\psi_k$. We have
\begin{align*}
&\E{\E{\left\|\sum_{k\in \cC_2} \psi_k-\sum_{k\in \cC_2}\tilde{g}_{k}\right\|^2|\mathbf{R}}}\\&=\sum_{k\in \cC_2}\E{\E{\|\psi_k-\tilde{g}_{k}\|^2|\mathbf{R}}}+ \sum_{k\neq k^\prime}\E{\E{\langle\psi_k-\tilde{g}_k, \psi_{k^\prime}-\tilde{g}_{k^\prime}\rangle | \mathbf{R}}}\\
&= \sum_{k\in \cC_2}\E{\|\psi_k-\tilde{g}_{k}\|^2}+ \sum_{k\neq k^\prime}\E{\langle\E{\psi_k-\tilde{g}_k|\mathbf{R}},\E{ \psi_{k^\prime}-\tilde{g}_{k^\prime}|\mathbf{R}}\rangle}\\
&=\sum_{k\in \cC_2}\E{\|\psi_k-\tilde{g}_{k}\|^2}+ \E{\left(\sum_{k\in \cC_2}\|\E{\psi_k-\tilde{g}_k|\mathbf{R}}\|\right)^2}-\E{\sum_{k\in \cC_2}\|\E{\psi_k-\tilde{g}_k|\mathbf{R}}\|^2}\\
&\leq \sum_{k\in \cC_2}\E{\|\psi_k-\tilde{g}_{k}\|^2}+\frac{K}{2}\sum_{k\in \cC_2}\E{\|\E{\psi_k-\tilde{g}_{k}|\mathbf{R}}\|^2},\numberthis \label{e:mse}
\end{align*}
where the second equality uses the fact that $\psi_k-\tilde{g}_{k}$ and $\psi_{k^\prime}-\tilde{g}_{k^\prime}$ are independent given $\mathbf{R}$, and 
the last line uses Jensen's inequality. Now
consider an event $\mathcal{V}_j=\{|\mathbf{R}S(j)|\leq M, |\mathbf{R}\tilde{g}_{k}(j)|\leq M\}$ and use unitary property of $\mathbf{R}$ to write the first term in \eqref{e:mse} as
\begin{align}\label{e:mse2}
    &\sum_{k\in \cC_2}\E{\|\mathbf{R}\psi_k-\mathbf{R}\tilde{g}_{k}\|^2}\\
    &=\sum_{k\in \cC_2}\sum_{j\in [d] }\E{(\mathbf{R}\psi_k(j)-\mathbf{R}\tilde{g}_{k}(j))^2\cdot \mathbbm{1}_{\mathcal{V}_j}} +\E{(\mathbf{R}\psi_k(j)-\mathbf{R}\tilde{g}_{k}(j))^2\cdot \mathbbm{1}_{\mathcal{V}_j^c}}
\end{align}
The first term in \eqref{e:mse2} is bounded as
\begin{align*}
\sum_{k\in \cC_2}\sum_{j\in [d] }\E{(\mathbf{R}\psi_k(j)-\mathbf{R}\tilde{g}_{k}(j))^2\cdot \mathbbm{1}_{\mathcal{V}_j}}
&\leq \sum_{k\in \cC_2}\E{\|\mathbf{R}\psi_k-\mathbf{R}\tilde{g}_{k}\|^2}\\
    &= \sum_{k\in \cC_2}\E{\E{\|\mathbf{R}\psi_k-\mathbf{R}\tilde{g}_{k}\|^2|S,\hat{g}_k}}\\
    &\leq \frac{2M\sqrt{d}}{I}\sum_{k\in \cC_2}\E{\|\mathbf{R}\tilde{g}_{k}-\mathbf{R}S\|}\\
    &\leq \frac{2M\sqrt{d}}{I}\sum_{k\in \cC_2}\sqrt{\E{\|\mathbf{R}\tilde{g}_{k}-\mathbf{R}S\|^2}}\\
    &\leq \frac{4M\sqrt{d}}{KI}\sqrt{\frac{4\sigma^2}{K}+\frac{32B^2d}{K(v-1)^2}+\sigma^2}\\
&\leq \frac{4\sqrt{3}M\sqrt{d}\sigma}{I},
\end{align*}
where the first inequality uses $\mathbbm{1}_{\mathcal{V}_j}\leq 1$,  the second inequality is due to Lemma \ref{l:daq}, the third inequality is Jensen's inequality, the fourth one uses the fact that $(K/2)S$ is the output of $\tt UQ-OTA$ along with the proof of Theorem \eqref{t:CUQ}, and the last inequality holds for parameters $v=7, K\geq B^2d/\sigma^2$.
For the second term in \eqref{e:mse2}, each summand is bounded as 
\begin{align*}
\E{(\mathbf{R}\psi_k(j)-\mathbf{R}\tilde{g}_{k}(j))^2\cdot \mathbbm{1}_{\mathcal{V}_j^c}}
    &\leq  8M^2\cdot \mathbb{P}(\mathcal{V}_j^c)+
2\E{(\mathbf{R}(S-\tilde{g}_{k})(j))^2\cdot \mathbbm{1}_{\mathcal{V}_j^c}}\\
&= 8M^2\cdot \mathbb{P}(\mathcal{V}_j^c)+
2\E{(\mathbf{R}(S-\tilde{g}_{k})(j))^2\cdot \mathbbm{1}_{\mathcal{V}_j^c}\cdot\mathbbm{1}_{\mathcal{V}_j^{\prime }}}\\
&\qquad+2\E{(\mathbf{R}(S-\tilde{g}_{k})(j))^2\cdot \mathbbm{1}_{\mathcal{V}_j^c}\cdot\mathbbm{1}_{\mathcal{V}_j^{\prime c}}}\\
&\leq 8M^2\cdot \mathbb{P}(\mathcal{V}_j^c)+2M^2\cdot\mathbb{P}(\mathcal{V}_j^c)+2\E{(\mathbf{R}(S-\tilde{g}_{k})(j))^2\cdot \mathbbm{1}_{\mathcal{V}_j^{\prime c}}}
\end{align*}
where the first inequality uses the definition of $\psi_k$ along with $(a+b)^2\leq 2(a^2+b^2)$, and the second equality follows from considering a new event $\mathcal{V}_j^{\prime}=\{|\mathbf{R}(S-\tilde{g}_k)(j)|\leq M\}$ and $\mathbbm{1}_{\mathcal{V}_j^{\prime c}}\leq 1$.
Note that for first two terms above $\mathbb{P}(\mathcal{V}_j^c)\leq 4e^{-\frac{dK^2M^2}{8B^2}}$. For the third term above, we note that $\mathbf{R}(S-\tilde{g}_{k})(j)$ is sub-Gaussian with variance factor $\frac{16B^2}{dK^2}$ (see for instance, \cite[Lemma V.8]{mayekar2020ratq}) and use the following concentration result. 
\begin{lem}\cite[Lemma 8.1]{mayekar2021wyner}\label{l:sG_conc}
For a sub-Gaussian random $Z$ with variance factor $\sigma^2$
and every $t\geq 0$, we have 
\[
\E{Z^2\indic{\{|Z|>t\}}}\leq 
2(2\sigma^2+t^2)e^{-t^2/2\sigma^2}.
\]
\end{lem}
\noindent Using $e^{-\frac{dK^2M^2}{8B^2}}\leq e^{-\frac{dK^2M^2}{32B^2}}$ and the Lemma \ref{l:sG_conc}, we now have
\begin{align*}
\sum_{k\in \cC_2}\sum_{j\in [d] }\E{(\mathbf{R}\psi_k(j)-\mathbf{R}\tilde{g}_{k}(j))^2\cdot \mathbbm{1}_{\mathcal{V}_j^c}}&\leq\left(\frac{64B^2}{K}+22M^2dK\right)e^{-\frac{dK^2M^2}{32B^2}}\\
    &\leq 26\delta^2,\numberthis\label{e:mse3}
\end{align*}
where for the last inequality, we choose $M^2{=}\frac{c_2B^2}{K^2d}\ln \left(\frac{c_2B}{\sqrt{K}\delta}\right)$, for $\delta\in \left(0,\frac{c_2B}{\sqrt{K}}\right)$. Further, the second term in \eqref{e:mse} is 0 under $\mathcal{V}$ and can be bounded under $\mathcal{V}^c$ using
the Jensen's inequality as 
\begin{align*}
 \sum_{k\in \cC_2}\E{\|\E{\psi_k-\tilde{g}_{k}|\mathbf{R}}\|^2}&= \sum_{k\in \cC_2}\sum_{j\in [d] }\E{(\E{\mathbf{R}\psi_k(j)-\mathbf{R}\tilde{g}_{k}(j)|\mathbf{R}})^2}\\
    &= \sum_{k\in \cC_2}\sum_{j\in [d] }\E{(\E{\mathbf{R}\psi_k(j)-\mathbf{R}\tilde{g}_{k}(j)|\mathbf{R}})^2\cdot \mathbbm{1}_{\mathcal{V}_j^c}}\\
    &\leq \sum_{k\in \cC_2}\sum_{j\in [d] }\E{\E{(\mathbf{R}\psi_k(j)-\mathbf{R}\tilde{g}_{k}(j))^2|\mathbf{R}}\cdot \mathbbm{1}_{\mathcal{V}_j^c}}\\
    &= \sum_{k\in \cC_2}\sum_{j\in [d] }\E{(\mathbf{R}\psi_k(j)-\mathbf{R}\tilde{g}_{k}(j))^2\cdot \mathbbm{1}_{\mathcal{V}_j^c}}\\
    &\leq 26\delta^2,
\end{align*}
where the first equality uses the unitary property of $\mathbf{R}$, the second uses the fact that boosted DAQ yields an unbiased estimate (see Lemma \ref{l:daq}), the first inequality follows from Jensen's inequality, and the last line uses the bound in \eqref{e:mse3}.
Finally, we set $\delta=\frac{c_2B}{K^2N}$.
Combining all above in \eqref{e:mse} and using \eqref{e:mean_square}, we get
\begin{align*}
\alpha^2(Q)
&= \frac{4\sqrt{3}M\sigma\sqrt{d}}{I}+26\delta^2+13K\delta^2+\frac{2\sigma^2}{K}\\
&\leq\frac{4\sqrt{3}B\sigma}{K}+\frac{3\sigma^2}{K},
\end{align*}
where the last line holds whenever $K\geq B^2d/\sigma^2$ and by choosing $I=\ceil{\sqrt{c_2\ln (\frac{c_2B}{\sqrt{K}\delta})}}$.
Also, 
\begin{align*}
\beta^2(Q) &= \|\E{\psi(Y)}-\sum_{k\in \cC_2}\tilde{g}_k\|^2\\
&\leq \left(\sum_{k\in \cC_2}\|\E{\mathbf{R}\psi_k-\mathbf{R}\tilde{g}_k}\|\right)^2\\
&\leq K \sum_{k\in \cC_2}\|\E{\mathbf{R}\psi_k-\mathbf{R}\tilde{g}_k}\|^2\\
&= K \sum_{k\in \cC_2}\sum_{j\in [d]}\E{(\mathbf{R}\psi_k(j)-\mathbf{R}\tilde{g}_k(j))\cdot \mathbbm{1}_{\mathcal{V}_j^c}}^2\\
&\leq K \sum_{k\in \cC_2}\sum_{j\in [d]}\E{(\mathbf{R}\psi_k(j)-\mathbf{R}\tilde{g}_k(j))^2\cdot \mathbbm{1}_{\mathcal{V}_j^c}}\\
&\leq 9K\delta^2\\
&= \frac{9c_5B^2}{K^3N^2},
\end{align*}
where the first identity follows from \eqref{e:unbiasedness}, the first inequality is triangle inequality, the second inequality uses Cauchy-Schwarz, the second identity uses unbiased nature of estimate under $\mathcal{V}_j$, the third inequality is Jensen's, and the last inequality uses the bound in \eqref{e:mse3}.
Since the final output is obtained in $d/p+d/p^\prime$ channel uses per iteration,
where $p=\left\lfloor\log_w\left(1+ \sqrt{\frac{K\SNR}{2\ln(KN^{1.5})}}\right)\right\rfloor$  and $p^\prime = \left\lfloor\log_{w^\prime}\left( 1+\sqrt{\frac{K\SNR}{2\ln(KN^{1.5})}}\right)\right\rfloor$ with $w=3K+1, w^\prime = KI/2+1$. 
Using  Lemma \ref{l:conv}, the proof is completed.
\section{Experiments}\label{s:exp}
\setcounter{figure}{0}   
  \pgfplotsset{compat=1.12, every axis/.append style={
  		line width=1pt, tick style={line width=0.6pt}}, width=7.2cm,height=5.5cm, tick label style={font=\tiny},
  	label style={font=\normalsize},
  	legend style={font=\normalsize}}
\begin{figure*}[ht]
\centering
\begin{minipage}{0.45\textwidth}
\centering
\begin{tikzpicture}
\begin{axis}[xmode=log,
ytick style={draw=none}, xtick style={draw=none}, legend style={nodes={scale=0.6}, font=\normalsize},
 	xlabel= $B/\sigma$,
 	ylabel= RMSE$\times \sqrt{\ell}$,
    legend pos= south east,
 	xmin=64/0.03,
 	xmax=8192/0.03,
 	ymin=0,
 	ymax=720,  grid=major, grid style={dotted, gray}
 	]
\addplot [mark=square, color=teal]
coordinates {
(2/0.03,  1.7539378) (4/0.03, 2.5105582) (8/0.03, 4.02280295) (16/0.03, 5.53571907) (32/0.03,  8.60321135) (64/0.03, 12.73694516) (128/0.03, 20.69575357) (256/0.03, 31.62981008) (512/0.03, 55.71783615) (1024/0.03, 96.240968) (2048/0.03,181.14406127) (4096/0.03, 367.24930002) (8192/0.03, 713.97056125)};
 	
\addplot [mark=otimes*, color=ta2gray]
table
[x expr=2*(2^\coordindex)/0.03, 
y expr=\thisrowno{0}
] {WZ_plots/WZ_50dB.dat}; 	

\legend{{\tt UQ-OTA}\\{\tt WZ-OTA}\\}
\end{axis}
\end{tikzpicture}
\vspace{-0.05cm}
\caption{Comparison of {\tt UQ-OTA} and {\tt WZ-OTA} at $\SNR=50$dB and $d=32$.}	
\label{fig1}
\end{minipage}
\hspace{0.05\textwidth}
\begin{minipage}{0.45\textwidth}
\centering
\begin{tikzpicture}
\begin{axis}[xmode = log, ytick style={draw=none}, xtick style={draw=none}, legend style={nodes={scale=0.6}, font=\normalsize},
 	xlabel= $B/\sigma$,
 	ylabel= RMSE$\times \sqrt{\ell}$,
    legend pos= south east,
 	xmin=64/0.03,
 	xmax=8192/0.03,
 	ymin=0,
 	ymax=710,  grid=major, grid style={dotted, gray}
 	]
\addplot [mark=square, color=teal]
table
[x expr=2*(2^\coordindex)/0.03, 
y expr=\thisrowno{0}
] {CUQ_plots/CUQ_75dB.dat}; 	
 	
\addplot [mark=otimes*, color=ta2gray]
table
[x expr=2*(2^\coordindex)/0.03, 
y expr=\thisrowno{0}
] {WZ_plots/WZ_75dB.dat}; 	

\legend{ {\tt UQ-OTA}\\ {\tt WZ-OTA}\\}
\end{axis}
\end{tikzpicture}
\vspace{-0.05cm}
\caption{Comparison of {\tt UQ-OTA} and {\tt WZ-OTA} at $\SNR=75$dB and $d=32$.}	
\label{fig2}
\end{minipage}
\end{figure*}

\begin{figure*}[ht]
\centering
\begin{minipage}{0.45\textwidth}
\centering
\begin{tikzpicture}
\begin{axis}[xmode = log,
ytick style={draw=none}, xtick style={draw=none}, legend style={nodes={scale=0.6}, font=\normalsize},
 	xlabel= $B/\sigma$,
 	ylabel= RMSE$\times \sqrt{\ell}$,
    legend pos= south east,
 	xmin=64/0.03,
 	xmax=4096/0.03,
 	ymin=0,
 	ymax=770,  grid=major, grid style={dotted, gray}
 	]
\addplot [mark=square, color=teal]
table
[x expr=2*(2^\coordindex)/0.03, 
y expr=\thisrowno{0}
] {CUQ_plots/CUQ_100dB.dat}; 	
 	
\addplot [mark=otimes*, color=ta2gray]
table
[x expr=2*(2^\coordindex)/0.03, 
y expr=\thisrowno{0}
] {WZ_plots/WZ_100dB.dat}; 	

\legend{{\tt UQ-OTA}\\{\tt WZ-OTA}\\}
\end{axis}
\end{tikzpicture}
\vspace{-0.05cm}
\caption{Comparison of {\tt UQ-OTA} and {\tt WZ-OTA} at $\SNR=100$dB and $d=64$.}	
\label{fig3}
\end{minipage}
\hspace{0.05\textwidth}
\begin{minipage}{0.45\textwidth}
\centering
\begin{tikzpicture}
\begin{axis}[xmode = log, ytick style={draw=none}, xtick style={draw=none}, legend style={nodes={scale=0.6}, font=\normalsize},
 	xlabel= $B/\sigma$,
 	ylabel= RMSE$\times \sqrt{\ell}$,
    legend pos= south east,
 	xmin=64/0.03,
 	xmax=4096/0.03,
 	ymin=0,
 	ymax=391,  grid=major, grid style={dotted, gray}
 	]
\addplot [mark=square, color=teal]
table
[x expr=2*(2^\coordindex)/0.03, 
y expr=\thisrowno{0}
] {CUQ_plots/CUQ_180dB.dat}; 	
 	
\addplot [mark=otimes*, color=ta2gray]
table
[x expr=2*(2^\coordindex)/0.03, 
y expr=\thisrowno{0}
] {WZ_plots/WZ_180dB.dat}; 	

\legend{ {\tt UQ-OTA}\\ {\tt WZ-OTA}\\}
\end{axis}
\end{tikzpicture}
\vspace{-0.05cm}
\caption{Comparison of {\tt UQ-OTA} and {\tt WZ-OTA} at $\SNR=180$dB and $d=64$.}	
\label{fig4}
\end{minipage}
\end{figure*}

\begin{figure*}[th]
\centering
\begin{minipage}{0.45\textwidth}
\centering
\begin{tikzpicture}
\begin{axis}[
ytick style={draw=none}, xtick style={draw=none}, legend style={nodes={scale=0.6}, font=\normalsize},
 	xlabel= $K$,
 	ylabel= RMSE$\times \sqrt{\ell}$,
    legend pos= north east,
 	xmin=200,
 	xmax=1000,
 	ymin=100,
 	ymax=500,  grid=major, grid style={dotted, gray}
 	]
\addplot [mark=square, color=teal]
table
[x expr=200*(\coordindex)+200, 
y expr=\thisrowno{0}
] {CUQ_plots/CUQ_100dB_K.dat}; 	
 	
\addplot [mark=otimes*, color=ta2gray]
table
[x expr=200*(\coordindex)+200, 
y expr=\thisrowno{0}
] {WZ_plots/WZ_100dB_K.dat}; 	

\legend{{\tt UQ-OTA}\\{\tt WZ-OTA}\\}
\end{axis}
\end{tikzpicture}
\vspace{-0.05cm}
\caption{Comparison of {\tt UQ-OTA} and {\tt WZ-OTA} at $\SNR=100$dB,  $B/\sigma=1.36\times 10^5$, and $d=64$.}	
\label{fig5}
\end{minipage}
\hspace{0.05\textwidth}
\begin{minipage}{0.45\textwidth}
\centering
\begin{tikzpicture}
\begin{axis}[ ytick style={draw=none}, xtick style={draw=none}, legend style={nodes={scale=0.6}, font=\normalsize},
 	xlabel= $K$,
 	ylabel= RMSE$\times \sqrt{\ell}$,
    legend pos= north east,
 	xmin=200,
 	xmax=2000,
 	ymin=100,
 	ymax=675,  grid=major, grid style={dotted, gray}
 	]
\addplot [mark=square, color=teal]
table
[x expr=200*(\coordindex)+200, 
y expr=\thisrowno{0}
] {CUQ_plots/CUQ_180dB_K.dat}; 	
 	
\addplot [mark=otimes*, color=ta2gray]
table
[x expr=200*(\coordindex)+200, 
y expr=\thisrowno{0}
] {WZ_plots/WZ_180dB_K.dat}; 	

\legend{ {\tt UQ-OTA}\\ {\tt WZ-OTA}\\}
\end{axis}
\end{tikzpicture}
\vspace{-0.05cm}
\caption{Comparison of {\tt UQ-OTA} and {\tt WZ-OTA} at $\SNR=180$dB, $B/\sigma=1.36\times 10^5$ ,  and $d=64$.}	
\label{fig6}
\end{minipage}
\end{figure*}
We demonstrate the performance of our proposed digital schemes {\tt UQ-OTA} and {\tt WZ-OTA} for the following mean estimation task under MAC constraints. 

 Each client $C_k$ has a $d$-dimensional vector $\hat{g}_k$, defined as $\hat{g}_k=\mu + U^{c}_k$, where $\mu\in [-1, 1]^d$ is a constant mean vector, and $U^{c}_k$ is a random vector whose coordinates are independently drawn from a uniform random variable denoted as ${\tt unif}(-\sigma^\prime, \sigma^\prime)$.
 Note that $\E{\hat{g}_k}=\mu$ and $\E{\|\hat{g}_k-\mu\|^2} = \frac{d\sigma^{\prime^2}}{3}$. The goal is to recover the sample average $\bar{g} = \frac{1}{K}\sum_{k\in K} \hat{g}_k$ which is an unbiased estimate for the mean vector $\mu$.
 
We compare the two proposed digital over-the-air schemes {\tt UQ-OTA} and {\tt WZ-OTA} for estimating $\bar{g}$.  We evaluate the performance of our proposed schemes by RMSE between the $\mu$ and the estimated sample average vector $\hat{\bar{g}}$ formed by the server.  We then plot a combined error metric which is the product of RMSE and the square root of number of MAC channel transmissions. Such a metric, in essence with the Lemma \ref{l:conv},  is a crucial term contributing to the overall performance analysis.  Our codes are implemented in Python language and are available online on GitHub~\cite{OTA_MAClink}.

We fix the number of clients $K=500$ and conduct the experiments for dimensions $d=32$ at $\SNR=50\text{dB and }75$dB, and for dimensions $d=64$ at $\SNR=100 \text{dB and }180$dB.  We fix the value of $\sigma^\prime =0.05196$ which gives a valid choice for $\sigma$ to be $0.03\sqrt{d}$. For all of these experiments,  we vary the $B/\sigma$ for different choices of $B$. All the experiments are averaged over 20 runs for statistical consistency.

In Figures \ref{fig1}, \ref{fig2}, \ref{fig3}, and \ref{fig4},  we observe that the {\tt WZ-OTA} outperforms the {\tt UQ-OTA} for large values of $B/\sigma$. However, for lower values of $B/\sigma$, {\tt UQ-OTA} performs better than {\tt WZ-OTA}. This observation is in accordance with the Remark \ref{r:wz_better}. 

In other direction, we note that for both the schemes, there is decrease in the error performance for the same $B/\sigma$ and dimension $d$ as the SNR increases. This is because of better channel decoding with increasing operating $\SNR$.

Next, we demonstrate the error performance of both the digital schemes for increasing number of clients.
Specifically, we fix the values of ratio $B/\sigma=1.36\times 10^5$ and the dimension $d=64$.  In Figure~\ref{fig5}, we show the error performances of our schemes at $\SNR=100$dB and for the number of clients $K=200, 400, 600, 800, \text{ and } 1000$.  In Figure~\ref{fig5}, we show the same at $\SNR=180$dB and for the number of clients $K=200, 400, 600, 800, 1000, 1200, 1400, 1600, 1800, \text{ and } 2000$.  As can be seen from both the figures, the error performance decreases with increase in the number of clients.

\section{Conclusion}\label{s:concl}
We provide an almost complete characterization of the min-max convergence rate of over-the-air distributed optimization. 
Our bounds show that a simple analog coding scheme is optimal at low values of $\SNR$, but they can be far from optimal at high values of $\SNR$ (Remark \ref{r:analog}). This observation mirrors the observation made by \cite{JMT22jsait}, albeit in the single client setting. Furthermore, we design an explicit digital communication scheme based on lattice coding to match our lower bound for all values of $\SNR$. We hope our work inspires other explicit communication schemes for similar distributed optimization problems. Our upper bound matches our lower bound up to a nominal $\sqrt{\log K + \log \log N}$ factor (Theorem \ref{t:UWZSGD}). Further closing the gap between our upper and lower bound would lead to new communication schemes or lower bound techniques for distributed optimization and is an exciting research direction.

\section*{Acknowledgement}
The author would like to thank Himanshu Tyagi, Prathamesh Mayekar, and Naina Nagpal for many useful discussions and help with problem formulation and proof ideas.
\bibliography{tit2018}
\bibliographystyle{ieeetr}
\appendix
\section*{Mathematical details concerning Remarks \ref{rem:uqota} and \ref{r:wz_better}}
\paragraph{Sub-optimality of {\tt UQ-OTA} at high SNR}
\noindent 
Let $x=\SNR,y=\frac{K}{2\ln(KN^{1.5})}$.  Then, we can write 
\begin{align*}
p&\geq  \frac{\log(1+\sqrt{xy})}{\log(Kd)}-1\\
&= \frac{\log\left(1+\sqrt{x}\right)}{{\log(Kd)}}+\frac{\log\left(\frac{1}{1+\sqrt{x}}+\frac{\sqrt{xy}}{1+\sqrt{x}}\right)}{\log(Kd)}-1\\
&\geq \frac{\log\left(1+\sqrt{x}\right)}{{\log(Kd)}}+\frac{\log\left(\min(1,\sqrt{y})\right)}{\log(Kd)}-1\\
&\geq \frac{\log\left(1+\sqrt{x}\right)}{{\log(Kd)}}-1\\
&\geq \frac{\frac{1}{2}\log\left(1+x\right)}{{\log(Kd)}}-1,
\end{align*}
where the first inequality holds due to the fact that $\floor{p}\geq p-1$, the second inequality is due to the term inside the second convex combination of points 1 and $\sqrt{y}$, the third inequality holds for $K$ sufficiently large satisfying $y\geq 1$, and the last one holds as $\log(1+\sqrt{x})=\frac{1}{2}\log(1+x+2\sqrt{x})\geq \frac{1}{2}\log(1+x)$.

\paragraph{Approximation of $p, p^{\prime}$ at large $K,N$ and $\SNR$}
At large $N$, we have
$w\leq w^\prime$ which implies $p\geq p^{\prime}$ and thus $q\geq p^{\prime}/2$.
Again, considering $x=\SNR,y=\frac{K}{2\ln(KN^{1.5})}$ and proceeding as earlier, we can show that 
$p^\prime \approx \frac{\frac{1}{2}\log(1+\SNR)}{\log(KI)}$ for large $\SNR\geq 2(2^d-1)$ regime.
\end{document}